\documentclass[11pt]{article}

\usepackage{latexsym,amssymb,fullpage}
\usepackage{amsmath,amscd}

\def\setminus{\smallsetminus}
\def\ens#1{\{#1\}}
\def\ie{\textit{i.e. }}
\def\eg{\textit{e.g. }}
\def\etal{\textit{et.al. }}
\def\al{\alpha}

\def\mcl{\mathcal} 
\newtheorem{prop}{Proposition}\def\PRO{\begin{prop}}\def\ORP{\end{prop}}
\newtheorem{coro}{Corollary}\def\COR{\begin{coro}}\def\ROC{\end{coro}}
\newtheorem{theo}{Theorem}\def\TH{\begin{theo}}\def\HT{\end{theo}}
\newtheorem{fact}{Fact}\def\FA{\begin{fact}}\def\AF{\end{fact}}
\newtheorem{prob}{Problem}\def\PRB{\begin{prob}}\def\BRP{\end{prob}}
\newtheorem{defi}{Definition}\def\DE{\begin{defi}}\def\ED{\end{defi}}
\newtheorem{lemme}{Lemma}\def\LE{\begin{lemme}}\def\EL{\end{lemme}}
\newcommand{\AR}[2][c]{$$\begin{array}[#1]{lllllllllllllll}#2\end{array}$$}

\def\EQ#1{\begin{eqnarray}#1\end{eqnarray}}

\def\ket#1{{|}#1\rangle}
\def\bra#1{\langle#1{|}}
\newcommand{\braket}[2]{\langle#1{|}#2\rangle}
\def\Prob{{\rm Prob}}
 
\newcommand{\norm}[1]{\Vert #1 \Vert}
\newcommand{\inp}[2]{({#1}\;,\;{#2})}
\def\01{\{0,1\}}
\newenvironment{proof}
{\noindent {\bf Proof. }}
{{\hfill $\Box$}\\
 \smallskip}
\newcommand{\COMMENT}[1]{}

\title{{Statistical Zero Knowledge and quantum one-way functions}}

\author{
Elham Kashefi\\
Christ Church, University of Oxford \&\\
IQC, University of Waterloo\\
elham.kashefi@comlab.ox.ac.uk
\and
Iordanis Kerenidis\\ 
University of Paris (LRI), CNRS \&\\
Dept. of Mathematics, M.I.T.\\
jkeren@lri.fr 
}

\begin{document}
\maketitle

\begin{abstract}
One-way functions are a fundamental notion in cryptography, since they are the necessary condition for the existence of secure encryption schemes. Most examples of such functions, including Factoring, Discrete Logarithm or the RSA function, can be, however, inverted with the help of a quantum computer. Hence, it is very important to study the possibility of {\em quantum one-way functions}, i.e. functions which are easily computable by a classical algorithm but are hard to invert even by a quantum adversary. 
In this paper, we provide a set of problems that are good candidates for quantum one-way functions. These problems include Graph Non-Isomorphism, Approximate Closest Lattice Vector and Group Non-Membership. More generally, we show that any hard instance of Circuit Quantum Sampling gives rise to a quantum one-way function. By the work of Aharonov and Ta-Shma \cite{AT02}, this implies that any language in Statistical Zero Knowledge which is hard-on-average for quantum computers, leads to a quantum one-way function. Moreover, extending the result of Impagliazzo and Luby \cite{IL89} to the quantum setting, we prove that quantum distributionally one-way functions are equivalent to quantum one-way functions.  
\end{abstract}
\pagebreak


\section{Introduction}

One-way functions are at the core of modern cryptography. The
fundamental task of cryptography is that of secure encryption of
information against malicious parties. The existence of such secure
encryption schemes implies that there is an efficient way of generating
instances of problems together with some auxiliary information, such
that it is easy to solve these instances with the help of the
auxiliary information but hard to solve on average without it.

This concept is exactly captured by the definition of one-way
functions, which are the necessary condition for the existence of
cryptography. Moreover, one-way functions have many theoretical
applications, for example in their connections to cryptographic
primitives like bit commitment and oblivious transfer, Zero Knowledge
Proof Systems and pseudorandom generators.

However, proving that one-way functions exist would
imply that $\rm{P} \neq \rm{NP}$ and hence, we only have ``candidate'' one-way 
functions. Such candidate problems include Factoring, Discrete
Logarithm, Graph Isomorphism, Quadratic Residuosity, Approximate Shortest Vector and Closest Vector and the RSA function. These 
problems seem to belong to a class called NP-\emph{Intermediate}, i.e. they
are NP problems for which we do not know any efficient algorithm,
but they don't seem to be NP-hard. Moreover, many of the candidate problems belong to the class of
Statistical Zero Knowledge (SZK). In fact, Ostrovsky \cite{Ost91}
showed that if SZK contains any {\em hard-on-average} problem, then
one-way functions exist.     

The emergence of quantum computation and communication has provided
the field of cryptography with many new strengths and challenges. The
possibility of unconditionally secure key distribution shows that
the laws of quantum mechanics can allow for the secure transmission of
information over quantum channels. Moreover, Shor's celebrated
algorithm for Factoring and Discrete Logarithm implies that many
classical one-way functions and hence cryptosystems, including RSA,
will not be secure against quantum adversaries. It is a very important
question to ask whether we can construct cryptosystems which are
secure even against quantum attacks. To this end, we need to find good
candidates for quantum one-way functions, i.e. functions which are
easily computable by a classical algorithm but hard to
invert even by a quantum adversary.  

Several other applications of quantum one-way functions have also been
studied in a series of papers. For example, the connections
between quantum one-way functions and quantum computationally secure
bit commitment schemes were explored in \cite{DMS00, AC01,CLS01}. 
On the other hand, Gottesman \etal  \cite{GC01} proposed a
digital signature scheme based on a quantum one-way function with
classical inputs but quantum outputs and proved the informational
security of their protocol. 
Moreover, Kashefi \etal
\cite{KNV02} and Kawachi \etal \cite{KKKP04} presented a necessary and
sufficient condition for testing the one-wayness of a given
permutation in the quantum setting based on the efficiency of
constructing a family of reflection operators.   
Recently, Watrous \cite{W05} proved that several classical
interactive proof systems are statistically zero-knowledge against
quantum attacks and showed that Computational Zero
Knowledge against quantum attacks for NP is implied by the   
existence of quantum one-way permutations.

Despite the importance of the applications of quantum one-way
functions, there had been few results so far that provided good candidate
problems \cite{DFS04}. Here, we prove the quantum analogue of Ostrovsky's result and
show that if there exists a problem in Statistical Zero Knowledge which is
hard-on-average for a quantum computer, then quantum one-way functions
exist and hence provide a set of problems that are
good candidates for quantum one-way functions. 

The key insight in our result is the connection of quantum one-way functions to the problem of {\em Circuit Quantum Sampling}.  
Informally speaking, \emph{quantum sampling} is the ability to prepare
efficiently a superposition that corresponds to a
samplable classical probability distributions, i.e. a superposition
whose amplitudes are the square roots of the probabilities of a
classical distribution from which one can efficiently sample. The
hardness of this task depends on the structure of the underlying
set. For example, it is well known that 
being able to quantumly sample from the set of homomorphisms of a
given input graph is sufficient to solve the notorious Graph
Isomorphism problem. Aharanov and Ta-shma \cite{AT02} have introduced
this framework of circuit quantum sampling and have shown that many problems
in quantum computation, including Graph Isomorphism, Discrete
Logarithm, Quadratic Residuosity and Approximate Closest Lattice Vector
(CVP), are all instances of it. 

We relate the problem of quantum sampling to quantum one-way functions by giving a simple proof that any hard instance of the quantum sampling problem implies the existence of a quantum one-way function. We first
prove our results for the case of one-to-one one-way functions, the 
existence of which seems to be a stronger assumption than that of
general one-way functions. Then, 
we generalize our results for many-to-one one-way
functions. We show that a hard instance of the CQS problem implies a quantum
distributionally one-way function  and then prove that a quantum
distributionally one-way function implies a quantum one-way function.  
The notion of classical distributionally one-way function was
introduced by Impagliazzo and Luby in \cite{IL89}, where they also
prove its equivalence to classical one-way function. 

Aharonov and Ta-Shma showed that any Statistical Zero
Knowledge language (SZK) can be reduced to a family of instances of
the CQS problem. Using our result that a hard instance of CQS implies the existence of a quantum one-way function, we conclude that if there exists a
language in Statistical  Zero Knowledge which is hard-on-average, then
quantum one-way functions exist.


\section{Preliminaries}

In this section we provide a brief overview of classical one-way functions and quantum computation. For an excellent exposition on quantum computation we refer the reader to \cite{NC00} and for one-way functions to \cite{Goldreich95}.  

\subsection{Classical one-way functions} 

\DE \label{d-c1way}
A function $f:\ens{0,1}^*\rightarrow \ens{0,1}^*$  is a \emph{weak one-way
function}, if the following conditions are satisfied:  
\begin {itemize}
\item[(i)] {\rm easy to compute:} $f$ can be computed by a polynomial
size classical circuit. 
\item[(ii)] {\rm slightly-hard to invert:} There exists a polynomial $p(\cdot)$
such that for any probabilistic polynomial time algorithm $I$ and for
all sufficiently large $n\in{\bf N}$ we have 
\[
\frac{1}{2^n}\sum_{x\in\{0,1\}^n} 
\Prob [ I(f(x),1^n)\in f^{-1}(f(x)) ] \leq 1- \frac{1}{p(n)}\,.
\]
\end{itemize}
\ED 

A classical weak one-way function $f$ is defined in terms of a uniform family of functions $f_n$, one for each input length $n$. The inverter $I$ of the function takes as input the value $f(x)$ and the size $n$ in unary. For simplicity, in the following definitions we omit the parameter $n$.
One can also assume, without loss of generality that the function 
$f$,  is \emph{length regular}  \ie for every $x,y \in \{0,1\}^*$,
if $|x|=|y|$ then $|f(x)|=|f(y)|$ and \emph{length preserving}  \ie
for every $x\in \{0,1\}^*$, $|f(x)|=|x|$ (for proof see \cite{Goldreich95}).

Intuitively, the above definition of a weak one-way function says that the function is easy to compute but the
probability that any algorithm fails to invert it, is not
negligible as Condition (ii) can be equivalently written in the following form:
\[
\frac{1}{2^n}\sum_{x\in\{0,1\}^n} 
\Prob [ I(f(x),1^n)\not\in f^{-1}(f(x)) ] \geq \frac{1}{p(n)}\,.
\]
Of course, such a definition seems to be very weak. One can define another type of one-way function, called \emph{strong one-way function}, where we require
that any algorithm inverts the function with negligible
probability, where Condition (ii) will be replaced as follows:
\[
\frac{1}{2^n}\sum_{x\in\{0,1\}^n} 
\Prob [ I(f(x),1^n) \in f^{-1}(f(x)) ] \leq \frac{1}{p(n)}\,.
\]
However, the two definitions are known to be equivalent
both in the classical and quantum setting \cite{Goldreich95,GS84,KNV02}, meaning that if a weak one-way function exists then a strong one-way function also exists. Hence, it suffices to work with the weaker but equivalent notion of weak one-wayness given in Definition \ref{d-c1way}.

Furthermore, Impagliazzo and Luby \cite{IL89} defined a seemingly weaker notion of one-wayness for many-to-one functions, called {\em distributionally one-way function}, and proved that, in fact, the existence of a distributionally one-way
function implies the existence of a one-way function.

\DE \label{d-cdist1way}
A function $f:\ens{0,1}^*\rightarrow \ens{0,1}^*$  is a \emph{distributionally one-way
function}, if the following conditions are satisfied:  
\begin {itemize}
\item[(i)] {\rm easy to compute:} $f$ can be computed by a polynomial
size classical circuit. 
\item[(ii)] {\rm hard to sample:} There exists a polynomial $p(\cdot)$
such that for any probabilistic polynomial time algorithm $S$ and for
all sufficiently large $n\in{\bf N}$, the distribution defined by $(x,f(x))$ and the distribution defined by $(S(f(x)),f(x))$ are statistically distinguishable by (i.e. have total variation distance) at least $\frac{1}{p(n)}$ when $x\in\ens{0,1}^n$ is chosen uniformly.
\end{itemize}
\ED 

\subsection{Quantum Computation}
Let $H$ denote a 2-dimensional complex vector space,
equipped with the standard inner product. We pick an orthonormal 
basis for this space, label the two basis vectors 
$\ket{0}$ and $\ket{1}$, and for simplicity identify them 
with the vectors $\left(\begin{array}{c}1\\ 0\end{array}\right)$
and $\left(\begin{array}{c}0\\ 1\end{array}\right)$, respectively.
A \emph{qubit} is a unit length vector in this space, and so can be
expressed as a linear combination of the basis states:
$$
\alpha_0\ket{0}+\alpha_1\ket{1}=\left(\begin{array}{c}\alpha_0\\
\alpha_1 \end{array}\right). 
$$
Here $\alpha_0,\alpha_1$ are complex \emph{amplitudes}, 
and $|\alpha_0|^2+|\alpha_1|^2=1$. 

An \emph{$m$-qubit system} is a unit vector in the $m$-fold tensor space
$H\otimes\cdots\otimes H$.  The $2^m$ basis states of this space are 
the $m$-fold tensor products of the states $\ket{0}$ and $\ket{1}$.
For example, the basis states of a 2-qubit system are the 4-dimensional 
unit vectors $\ket{0}\otimes\ket{0}$, $\ket{0}\otimes\ket{1}$,
$\ket{1}\otimes\ket{0}$, and $\ket{1}\otimes\ket{1}$.
We abbreviate, \eg, $\ket{1}\otimes\ket{0}$ to $\ket{0}\ket{1}$, or 
$\ket{1,0}$, or $\ket{10}$, or even $\ket{2}$ (since 2 is 10 in binary).
With these basis states, an $m$-qubit state $\ket{\phi}$ 
is a $2^m$-dimensional complex unit vector
$$
\ket{\phi}=\sum_{i\in\01^m}\alpha_i\ket{i}.
$$
We use $\bra{\phi}=\ket{\phi}^*$ to denote the conjugate transpose
of the vector $\ket{\phi}$, and $\inp{\phi}{\psi}=\bra{\phi}\cdot\ket{\psi}$
for the inner product between states $\ket{\phi}$ and $\ket{\psi}$.
These two states are \emph{orthogonal} if $\inp{\phi}{\psi}=0$.
The \emph{norm} of $\ket{\phi}$ is $\norm{\phi}=\sqrt{|\inp{\phi}{\phi}|}$. 

A quantum state can evolve by a unitary operation or by a
measurement. A \emph{unitary} transformation is a linear mapping that
preserves the $\ell_2$ norm. If we apply a unitary $U$ to a state
$\ket{\phi}$, it evolves to $U\ket{\phi}$.

The most
general measurement allowed by quantum mechanics is 
specified by a family of positive semidefinite operators 
$E_i=M_i^*M_i$, $1\leq i\leq k$, subject to the condition
that $\sum_i E_i=I$. A projective measurement is defined in the
special case where the operators are projections. 
Let $\ket{\phi}$ be an $m$-qubit state and
$B=\{\ket{b_1},\ldots,\ket{b_{2^m}}\}$ an orthonormal basis
of the $m$-qubit space. A projective measurement of the state
$\ket{\phi}$ in the $B$ basis means that we apply the projection operators
$P_i=\ket{b_i}\bra{b_i}$ to $\ket{\phi}$. The resulting quantum state
is $\ket{b_i}$ with probability $p_i=|\inp{\phi}{b_i}|^2$. 
 


\subsection{Quantum Sampling}
 
Let $\{C_i\}$ be a uniform classical circuit family and for every input size $n$ define $D_{C_n}$ to be the distribution over outputs of the circuit
$C_n:\01^n\rightarrow \01^m$ when the input distribution is uniform.  Denote by
$\ket{C_n} =
\sum_{z\in\{0,1\}^m}\sqrt{D_{C_n}(z)}\ket{z}$, the \emph{quantum
sample of outputs} of $C_n$. 

\DE Given a uniform family of classical circuit $\{C_i\}$ and a real number $0\leq \epsilon< \frac{1}{2}$, define $QS_C$ to be an efficient quantum circuit which for any sufficiently large input size $n$, prepares a state that is
$\epsilon$-close to the quantum sample $\ket{C_n}$, i.e.  
$
| \inp{QS_C(\ket{0},1^n)}{\ket{C_n}}|^2 \geq 1-\epsilon.
$
\ED

The problem of finding such a quantum circuit $QS_C$ for any given uniform family of classical circuits $\{C_i\}$ was
introduced by Aharanov and Ta-shma in \cite{AT02}, as the
\emph{Circuit Quantum Sampling Problem} (CQS). In fact, they
defined CQS as $\norm{QS_C(\ket{0},1^n)-\ket{C_n}}\leq \epsilon$, however both
definitions suffice for the proof that Statistical Zero Knowledge
reduces to a family of instances of the CQS problem. We say that the
quantum sampling problem for $\{C_i\}$ is {\em hard} if there
exists no efficient $QS$ for any constant $\epsilon\in
[0,1/2]$.

\section{Definitions of quantum one-way functions}

A {\em quantum one-way function} is defined similarly to the classical
case, where now the inverter $I$ is a polynomial size uniform quantum circuit family. For simplicity, we follow again the convention of omitting the parameter of the input size $n$.   

\DE \label{d-q1way}A one-to-one function $f:\ens{0,1}^*\rightarrow
\ens{0,1}^*$ is a \emph{weak quantum one-way function}, if the following
conditions are satisfied:  
\begin {itemize}
\item[(i)] {\rm easy to compute:} $f$ can be computed by a polynomial
size classical circuit. 
\item[(ii)] {\rm slightly-hard to invert:} There exists a polynomial $p(\cdot)$
such that for any quantum polynomial time algorithm $I$ and all
sufficiently large $n\in{\bf N}$ we have 
\[
\frac{1}{2^n}\sum_{x\in\{0,1\}^n} 
\Prob [ I(f(x))\in f^{-1}(f(x)) ] \leq 1- \frac{1}{p(n)}\,.
\]
\end{itemize}
\ED

In the quantum case, the probability of success of the inverter $I$ is
defined as the square of the inner product between the outcome of $I$
and the outcome of the perfect inverter $P$, where 
\[ P: \ket{f(x)}\ket{\beta}  \mapsto \ket{f(x)}\ket{x\oplus \beta}\,. \] 
In other words, for the case of one-to-one functions 
\[ \Prob [ I(f(x))\in f^{-1}(f(x)) ] =  \Prob [ I(f(x))= x ] =
|\inp{I(\ket{f(x)}\ket{\beta})}{\ket{f(x)}\ket{x\oplus \beta}}|^2.\]

As said before, one can also define another type of quantum 
one-way function (strong quantum one-way function), where we require
that any quantum algorithm inverts the function with
negligible probability (instead of just failing with non-negligible probability). However, similar to the classical case, if there exists a weak quantum one-way function (Definition \ref{d-q1way}), then there exists a strong quantum one-way function as well \cite{Goldreich95,GS84,KNV02}. In this article, one-way function means a weak one-way function if not stated otherwise.  

We now provide an alternative definition for a one-to-one quantum one-way function, which is more suitable for constructing the relation between  quantum one-way functions and the CQS problems and prove the equivalence of the two definitions. 

\DE
\label{d-q1way2} A one-to-one function $f:\ens{0,1}^*\rightarrow
\ens{0,1}^*$ is a {\em weak quantum one-way function} if:  
\begin{itemize}
\item[(i)] $f$ can be computed by a polynomial size classical circuit.
\item[(ii)] There exists a polynomial $p(\cdot)$ such that there
exists no quantum polynomial time algorithm $I'$ with the property
that for all sufficiently large $n\in{\bf N}$ we obtain 
\EQ{
I' : \ket{f(x)}\ket{\beta}\mapsto a_{f(x)} \ket{f(x)}\ket{x\oplus \beta} + b_{f(x)} \ket{f(x)}\ket{G_{f(x)}} \,,
}
where $G_{f(x)}$ is a garbage state,
$\frac{1}{2^n}\sum_{x\in\{0,1\}^n} a_{f(x)}^2 \geq 
1 - \frac{1}{p(n)}$ and $a_{f(x)}$ are positive real numbers.   
\end{itemize}
\ED

It is clear that definition \ref{d-q1way} implies definition
\ref{d-q1way2} and we also prove the converse.

\begin{theo}\label{theo-one}
If a one-to-one function $f$ is weak quantum one-way according to
definition \ref{d-q1way2}, then it is also weak quantum one-way according to definition \ref{d-q1way}.     
\end{theo}

\begin{proof}
Let $f:\01^*\rightarrow \01^*$ be a quantum one-way function according
to definition \ref{d-q1way2}. Assume for contradiction that this
function is not one-way according to definition \ref{d-q1way}. Then,
for all polynomials $p(\cdot)$ there exists a quantum
polynomial time algorithm $I$ with the property that for all
sufficiently large $n\in{\bf N}$  
\[
\frac{1}{2^n}\sum_{x\in\{0,1\}^n} 
\Prob [ I(f(x))\in f^{-1}(f(x)) ] \geq 1- \frac{1}{p(n)}\,,
\] 
or equivalently 
\EQ{
I : \ket{f(x)}\ket{\beta}\mapsto c_{f(x)} \ket{f(x)}\ket{x\oplus \beta} + d_{f(x)}\ket{\psi_{f(x)}}\,,  
}
where $\ket{\psi_{f(x)}}$ is a garbage state and
$\frac{1}{2^n}\sum_{x\in\{0,1\}^n} |c_{f(x)}|^2 \geq  
1 - \frac{1}{p(n)}$. Without loss of generality we can assume that
$c_{f(x)}$ are real numbers since it is well known that any quantum circuit with complex amplitudes can be replaced by another circuit with one more qubit and real amplitudes.     
We use this inverter to construct the following unitary that achieves the positive amplitudes. For clarity, here and in subsequent places in the paper we only show the unitary construction for the case where the ancilla registers are set to $\ket{0}$, unless the general ancilla state is required for the construction. It is clear of course how to unitarily extend the $\ket{0}$ ancilla to the other basis states.
\AR{
\ket{f(x)}\ket{0}\ket{0}\ket{0}
&\rightarrow_{(\rm{CNOT})_{1,3}}&
\ket{f(x)}\ket{0}\ket{f(x)}\ket{0}\\
&\rightarrow_{I_{1,2}}&(c_{f(x)}\ket{f(x)}\ket{x}+
d_{f(x)}\ket{\psi_{f(x)}})\ket{f(x)}\ket{0}\\  
&\rightarrow_{I_{3,4}}&c^2_{f(x)}\ket{f(x)}\ket{x}\ket{f(x)}\ket{x}+
c_{f(x)}d_{f(x)}\ket{f(x)}\ket{x}\ket{\psi_{f(x)}}+\\ 
&& d_{f(x)}c_{f(x)}\ket{\psi_{f(x)}}\ket{f(x)}\ket{x}+
d^2_{f(x)}\ket{\psi_{f(x)}}\ket{\psi_{f(x)}}\\ 
&\rightarrow_{(\rm{CNOT})_{1,3}(\rm{CNOT})_{2,4}}&c^2_{f(x)}\ket{f(x)}\ket{x}\ket{0}\ket{0}+
b_{f(x)}\ket{\psi^{\prime}_{f(x)}}\,,
}
where $\ket{\psi_{f(x)}^{\prime}}$ is the new garbage state, orthogonal to the
ideal state $\ket{f(x)}\ket{x}\ket{0}\ket{0}$ and by the fact that the
average of the squares is larger than the square of the average we have
\AR{\frac{1}{2^n}\sum_{x\in\{0,1\}^n} c^4_{f(x)} \geq
( \frac{1}{2^n}\sum_{x\in\{0,1\}^n} c^2_{f(x)} )^2 
\geq (1 - \frac{1}{p(n)})^2 \geq 1 - \frac{1}{p'(n)}\,.
}
Hence we have a new inverter
\EQ{ 
I': \ket{f(x)}\ket{\beta} \mapsto a_{f(x)} \ket{f(x)}\ket{x\oplus \beta} +
b_{f(x)} \ket{\psi_{f(x)}^{\prime}}\,, 
}
with $\frac{1}{2^n}\sum_{x\in\{0,1\}^n} a^2_{f(x)} \geq
1 - \frac{1}{p(n)}$ and $a_{f(x)} = c_{f(x)}^2$ being positive real numbers.
Finally, we can obtain the required form of the garbage state: 
\AR{
\ket{f(x)}\ket{0}\ket{0}&\rightarrow_{(\rm{CNOT})_{1,2}}&
\ket{f(x)}\ket{f(x)}\ket{0}\\ 
&\rightarrow_{I'_{2,3}}& a_{f(x)}\ket{f(x)}\ket{f(x)}\ket{x}+
b_{f(x)}\ket{f(x)}\ket{\psi_{f(x)}^{\prime}}\\  
&\rightarrow_{(\rm{CNOT})_{1,2}}&a_{f(x)}\ket{f(x)}\ket{0}\ket{x}+
b_{f(x)}\ket{f(x)}\ket{G_{f(x)}}\,.
}
We reached a contradiction and therefore the function $f$ is one-way according to definition \ref{d-q1way}. 
Note that for simplicity of presentation we dropped the $\ket{0}$ registers that are constant for all $x$.
\end{proof}

The important aspect of Theorem \ref{theo-one} is the positivity of the amplitude $a_{f(x)}$ in the definition of the inverter algorithm $I'$. We will use this fact in order to relate one-way functions and circuit quantum sampling.  

In the standard definition, a many-to-one function is called one-way
if there exists no inverter that outputs with high probability an arbitrary
preimage of $f(x)$. For many-to-one functions, Impagliazzo and Luby \cite{IL89} 
defined a seemingly weaker notion, the {\em distributionally one-way
function}. In this case, an inverter is required to output a {\em
random} preimage of $f(x)$ and not just an arbitrary one. However,
they prove that, in fact, the existence of a distributionally one-way
function implies the existence of a one-way function. We also define quantum distributionally one-wayness for many-to-one functions and will prove its equivalence to the quantum one-way functions.

\DE \label{d-dist1way}A many-to-one function $f:\ens{0,1}^*\rightarrow
\ens{0,1}^*$ is a
\emph{quantum distributionally one-way function}, if the following
conditions are satisfied: 
\begin {itemize}
\item[(i)] $f$ can be computed by a polynomial size classical circuit.
\item[(ii)] {\rm hard to invert:} There exists a polynomial $p(\cdot)$
such that for any quantum polynomial time algorithm $S$ and all
sufficiently large $n\in{\bf N}$ we have
\[
\frac{1}{2^n}\sum_{x\in\{0,1\}^n} 
| \inp{S(\ket{f(x)}\ket{0})} {\ket{f(x)}\ket{H_{f(x)}}}|^2 \leq
1-\frac{1}{p(n)}, 
\]
where $\ket{H_{f(x)}} = \frac{1}{\sqrt{|f^{-1}(f(x))|}}\sum_{x\in
f^{-1}(f(x))}\ket{x}$. 
\end{itemize}
\ED
Note that one could potentially consider different definitions for
quantum distributionally one-way functions, for example the quantum
inverter could return a superposition with equal amplitudes but
different phases. We believe that our quantum definition captures the
essence of the classical one and moreover, we only use the 
above notion as an intermediate step in our proofs. Similar to the
case of one-to-one functions we also give an equivalent definition 

\DE
\label{d-dist1way2}
A many-to-one function $f:\ens{0,1}^*\rightarrow \ens{0,1}^*$ is a
{\em quantum distributionally one-way function} if: 
\begin{itemize}
\item[(i)] $f$ can be computed by a polynomial size classical circuit.
\item[(ii)] There exists a polynomial $p$ such that there exists no
quantum polynomial time algorithm $S'$ with the property that for all
sufficiently large $n\in{\bf N}$ we obtain
\EQ{
S' : \ket{f(x)}\ket{0}\mapsto a_{f(x)}
\ket{f(x)}\ket{H_{f(x)}} + b_{f(x)} \ket{f(x)}\ket{G_{f(x)}}\,,
} 
where $\ket{G_{f(x)}}$ is a garbage state,
$\frac{1}{2^n}\sum_{x\in\{0,1\}^n} a^2_{f(x)}
\geq 1 - \frac{1}{p(n)}$, $a_{f(x)}$ are positive real numbers and
$\ket{H_{f(x)}}~=~\frac{1}{\sqrt{|f^{-1}(f(x))|}}\sum_{x\in 
f^{-1}(f(x))}\ket{x}$.
\end{itemize}
\ED
We can easily extend the above algorithm $S'$ into a unitary operation by mapping every other basis state $\ket{f(x)}\ket{\beta}$ to $\ket{f(x)}\ket{T^{\beta}_{f(x)}}$, where the set $\{\ket{H_{f(x)}}, T^1_{f(x)},\ldots,T^{2^n-1}_{f(x)}\}$ is any orthonormal basis.   
Following the same steps as in the proof of Theorem \ref{theo-one} we
have
 
\begin{theo}\label{theo-dist}
If a many-to-one function $f$ is quantum one-way according to
definition \ref{d-dist1way2}, then it is also quantum one-way according to definition \ref{d-dist1way}.        
\end{theo}


\section{Circuit quantum sampling and one-way functions}

In this section, we show that hard instances of the Circuit Quantum
Sampling problem are good candidates for quantum one-way functions.

\subsection{One-to-one one-way functions}\label{1-1}

We first focus our attention to the case of one-to-one one-way
functions. The existence of one-to-one one-way functions is a seemingly
stronger assumption than that of the existence of general one-way
functions, since a one-way function doesn't immediately imply a
one-to-one one-way function. However, this case illustrates the main
ideas of our construction. In the following 
sections, we generalize our results for the case of many-to-one
functions.  
  
\begin{theo} 
\label{t-qs} Assume for a classical circuit family $\{C_n\}$, which computes a
one-to-one function, the corresponding CQS problem is hard , \ie
there exists no efficient quantum circuit implementing
$QS_C$. Then the function $f:\01^* \rightarrow \01^*$ which is defined for every input size $n$ as $f_n: x \mapsto C_n(x)$ is a quantum one-way function. 
\end{theo}

\begin{proof} 
For clarity, we are going to omit the parameter of the input size $n$ from the inverter. Since, the circuit is  efficient, one can implement the unitary map 
\EQ{\label{e-f}
U_f : \ket{x}\ket{0} \mapsto \ket{x}\ket{f(x)}\,,
}

The theorem follows by proving the contrapositive. Assume that $f$ is
not a quantum one-way function. Then according to definition
\ref{d-q1way2}, for every   
polynomial $p$ there exists a quantum circuit $I'$ which succeeds in
approximately inverting $f$, i.e. for all sufficiently
large $n\in{\bf N}$ we have 
\EQ{\label{e-invers3}
I' : \ket{f(x)}\ket{\beta}\mapsto a_{f(x)} \ket{f(x)}\ket{x\oplus \beta}+b_{f(x)} \ket{f(x)}\ket{G_{f(x)}} \,,
}
where $\ket{G_{f(x)}}$ is a garbage state, 
$\frac{1}{2^n}\sum_{x}a^2_{f(x)} > 1 - \frac{1}{p(n)}$ and the
$a_{f(x)}$'s are positive. Now, 
from equations \ref{e-f} and \ref{e-invers3} we have 
\AR{
\ket{x}\ket{0}&\rightarrow_{U_f}& \ket{x}\ket{f(x)}\\
&\rightarrow_{\rm{SWAP}}& \ket{f(x)}\ket{x}\\
&\rightarrow_{I'}& a_{f(x)} \ket{f(x)}\ket{0}+b_{f(x)}
\ket{f(x)}\ket{G^{'}_{f(x)}}\,.
}

Starting with a uniform superposition of $x\in \01^n$ we have
\begin{eqnarray*}
\frac{1}{2^{n/2}}\sum_{x\in \01^n}\ket{x}\ket{0} & \rightarrow &
\frac{1}{2^{n/2}}\sum_{x} ( \; a_{f(x)}\ket{f(x)}\ket{0} + b_{f(x)} 
\ket{f(x)}\ket{G^{'}_{f(x)}} \;) \equiv \ket{Q_n} \,.
\end{eqnarray*}

We claim that the above circuit that on input $(\ket{0},1^n)$ outputs $\ket{Q_n}$ is a quantum sampler for $C$. Let $\ket{C_n} =  \frac{1}{2^{n/2}} \sum_{x}\ket{f(x)}\ket{0}$ be the quantum sample of the circuit $C$, then
\begin{eqnarray*}
| \braket{Q_n}{C_n} |^2 = | \frac{1}{2^n} \sum_{x}a_{f(x)}|^2 \geq  |
\frac{1}{2^n} \sum_{x}a^2_{f(x)}|^2 >  (1 - 1/p(n))^2 > 1-\epsilon \,,
\end{eqnarray*}
where $\epsilon = \frac{2}{p(n)}-\frac{1}{p^2(n)}$. This is a contradiction to $C$ being a hard instance of the CQS problem and hence $f$ is a quantum one-way function.
\end{proof}



\subsection{Many-to-one one-way functions}

The previous section dealt with the case of one-to-one one-way
functions. Here, we generalize our results to the case of
many-to-one functions. We show that the existence of a hard instance of CQS
problem, where the circuit family $\{C_n\}$ is many-to-one, implies the existence
of a quantum distributionally one-way function. In the next section we
prove that a quantum distributionally one-way function implies a
quantum one-way function.

\begin{theo} 
\label{t-distqs} Assume for a classical circuit family $\{C_n\}$, which computes a many-to-one function,  the corresponding 
CQS problem is hard , \ie there exists no efficient
quantum circuit implementing $QS_C$. Then the function $f:\01^*\rightarrow \01^*$ which is defined for every input size $n$ as
$f_n: x \mapsto C_n(x)$ is a quantum distributionally one-way function. 
\end{theo}

\begin{proof}
Since the classical circuit is efficient one can implement the unitary
map 
\[
U_f : \ket{x}\ket{0} \mapsto \ket{x}\ket{f(x)}\,.
\]
Assume that $f$ is not a quantum distributional one-way, then
according to definition \ref{d-dist1way2} for every
polynomial $p$ there exists a quantum polynomial time algorithm $S'$
which succeeds in 
approximately implementing a sampler for $f$, i.e. for all sufficiently
large $n\in{\bf N}$ we have 
\EQ{\label{e-d-invers3}
S' : \ket{f(x)}\ket{0}\mapsto a_{f(x)}
\ket{f(x)}\ket{H_{f(x)}} + b_{f(x)} \ket{f(x)}\ket{G_{f(x)}} \,,
}
where $\frac{1}{2^n}\sum_{x\in\{0,1\}^n} a^2_{f(x)} >
1 - \frac{1}{p(n)}$ and the $a_{f(x)}$'s are positive.
Note that one can unitarily extend the $S'$ to apply over any state of the form $\ket{f(x)}\ket{\beta}$ with $\beta \not = 0$.
Using the above unitaries, we can construct a quantum
sampler $QS_C$ that for every input $n$ constructs a quantum sample for $C_n$: 
\begin{eqnarray*}
\sum_{x\in\{0,1\}^n}\frac{1}{2^{n/2}}\ket{x} \ket{0} & \equiv &
\sum_{f(x)}\frac{\sqrt{|f^{-1}(f(x))|}}{2^{n/2}}\ket{H_{f(x)}}\ket{0}\\
& \rightarrow_{U_f}   &
\sum_{f(x)}\frac{\sqrt{|f^{-1}(f(x))|}}{2^{n/2}}\ket{H_{f(x)}}\ket{f(x)}\\ 
& \rightarrow_{\rm{SWAP}}& 
\sum_{f(x)} \frac{\sqrt{|f^{-1}(f(x))|}}{2^{n/2}} \ket{f(x)}\ket{H_{f(x)}}
\\
& \rightarrow_{S^{'\dagger}} & \sum_{f(x)} \frac{\sqrt{|f^{-1}(f(x))|}}{2^{n/2}}
(\; a_{f(x)}\ket{f(x)}\ket{0} + b_{f(x)} \ket{f(x)}\ket{G'_{f(x)}}) \equiv \ket{Q_n}\,.
\end{eqnarray*}

The quantum sample for the circuit $C_n$ is $ \ket{C_n}=\sum_{f(x)}
\frac{\sqrt{|f^{-1}(f(x))|}}{2^{n/2}} \ket{f(x)}\ket{0}$. Similarly to the
proof of Theorem \ref{t-qs}: 
\AR{
| \braket{Q_n}{C_n}|^2 &=& | \sum_{f(x)}\frac{|f^{-1}(f(x))|}{2^n}a_{f(x)}|^2 \\
  &=&  |\frac{1}{2^n}\sum_{x\in\01^n} a_{f(x)}|^2 \\
  &\geq&  |\frac{1}{2^n}\sum_{x\in\01^n} a^2_{f(x)}|^2 \\
  &>&  (1 - 1/p(n))^2 > 1-\epsilon\,,
}
where $\epsilon = \frac{2}{p(n)}-\frac{1}{p^2(n)}$. This is a contradiction and hence, $f$ is a quantum distributionally
one-way function.  
\end{proof}


\subsection{From quantum distributionally one-way functions to quantum
one-way functions}  

In the classical setting, Impagliazzo and Luby \cite{IL89} proved that
the existence of a distributionally one-way function implies the
existence of a one-way function. In this section, we
describe the main ideas of their construction and show how to
prove the equivalent result in the quantum setting. 

\TH \label{t-dist2oneway}
If there exists a quantum distributionally one-way function then there
exists a quantum one-way function. 
\HT

\subsubsection{The Impagliazzo and Luby construction}

Let $f:\01^*\rightarrow \01^*$ be a candidate distributionally one-way
function. Then, there exists a function $g$ such that an
inverter $I$ for $g$ implies the existence of a sampler $S$ for
$f$. Let us fix the size of input to $n$, this can be done as  we are working with a uniform  circuit family. 
More precisely, Impagliazzo and Luby showed that if there exists an inverter $I$ for $g$ that succeeds with probability $1- \delta^2/n$, then there exists a sampler $S$ for $f$, such that the distributions $(S(f(x)),f(x))$ and $(x,f(x))$ are $O(\delta)$-close in total variation distance ($\delta$ is the inverse of a large polynomial). Without loss of generality, the inverter for $g$ outputs $\perp$ when it's given as input something which is not in the image of $g$. 

Now, let us try to describe the main ideas of their construction. First,
assume that for a given $f(x)$ we know the size of the preimage
$|f^{-1}(f(x))|$ and let $k = \lfloor \log |f^{-1}(f(x))|\rfloor +
O(\log n)$. We define the function $g$ as  
\[ g(x,h_k) = (f(x),h_k,h_k(x)) \,.\]
In other words, $g$ takes as inputs an $x$ and a random
string $h_k$ which can be thought of as a random universal hash function
$h_k:\01^n\rightarrow \01^k$. The output of $g$ is the value $f(x)$,
the random universal hash function and the output of the hash function on $x$. 

There are two observations to be made about the hash
function. First, since the range of the hash function is slightly
larger than the number of $x$'s in the preimage of $f(x)$, with high
probability the mapping $x \mapsto h_k(x)$ for $\{x \in
f^{-1}(f(x))\}$ is a one-to-one mapping. This implies, that if we
could pick uniformly an element from the set $\{h_k(x)| x\in
f^{-1}(f(x))\}$ then the inverter of $g$ on input $(f(x),h_k,h_k(x))$
would return a uniform $x \in f^{-1}(f(x))$. 

Second, it's indeed possible to pick a uniform element of the set
$\{h_k(x)| x\in f^{-1}(f(x))\}$. Since the range
of the hash function is not too much larger than the size of the
preimage of $f(x)$, if we pick a random element $r_k \in \01^k$, then
with non negligible probability it holds that $r_k = h_k(x)$ for some $x\ \in
f^{-1}(f(x))$. By repeating the process a polynomial number of times, we can achieve high success probability. 

The above two properties enable one to prove that, when one knows the
size of the preimage of $f(x)$, the following
procedure is a sampler for $f(x)$:\\ 

\noindent
{\bf Partial Sampler PS(f(x),k)}\\
\hspace*{1em} Repeat a polynomial number of times  \\
\hspace*{3em}Pick a random hash function $h_k$ and $r_k\in \01^k$.\\
\hspace*{3em}If $I(f(x),h_k,r_k)\neq \perp$ then output it and exit.\\
\hspace*{1em} Output $\perp$ \,.\\

The remaining issue is that the sampler doesn't know the size of
the preimage of $f(x)$. Suppose we pick the range of the hash function
to be much larger than the actual size of the preimage of $f(x)$. Then
the above sampler outputs $\perp$ with very high probability. However,
conditioned on it producing an output $x$, then this $x$ is still
almost uniformly distributed in $\{f^{-1}(f(x))\}$. This is true since
the hash function randomly hashes $|f^{-1}(f(x))|$ values of $x$ to a
much larger range, and therefore, the mapping is with very high
probability one-to-one. 
 
Hence, we can construct a sampler for $f$ by starting with the largest
possible value for the range of the hash function and keep decreasing
it until there is an outcome:\\

\noindent
{\bf Sampler S(f(x))}\\
\hspace*{1em} For $j = n + O(\log n)$ to $O(\log n)$: \\
\hspace*{3em}If $PS(f(x),j)\neq \perp$ output it and exit.\\
\hspace*{1em} Output $\perp$\,.\\

Impaglazzo and Luby show that the overall errors  of the sampler $S$ are at most $O(\delta)$, i.e. inverse polynomially small. Their analysis is based on the following claims proved in \cite{IL89}:
\begin{enumerate} 
\item The errors from the fact that the hash function $h_k$ is not truly one-to-one are negligible for all values $j \geq k$.  
\item Since the inverter for $g$ is not perfect, the sampler doesn't work for every $f(x)$ but for $f(x)$'s that correspond to at least a $(1-\delta)$ fraction of the $x$'s (we call such $f(x)$ `good'). This is sufficient, since the total error from the rest of the inputs is at most $O(\delta)$. Moreover, for these `good' $f(x)$'s the inverter $I$ of $g$ succeeds with probability $(1-O(\delta))$. 
\item In the case of a `good' $f(x)$, if the sampler produces an
output for a $j\geq k$, then this $x$ is guaranteed to be almost uniform (i.e. the distributions $(S(f(x)),f(x))$ and $(x,f(x))$ have $O(\delta)$ total variation distance). 
\item In the case of a `good' $f(x)$, the probability
that the sampler actually produces an output for $j \geq k$ is, in fact,
very close to 1 (i.e. $1-O(\delta)$). 
\end{enumerate}
We will also need the following precise lemma from \cite{IL89}  
\begin{lemme}\cite{IL89}\label{pk}
Let $p_j$ be the probability that the Partial Sampler $PS(f(x),j)$ produces a
legal output. Then, for all $j \geq  k = \lfloor \log
|f^{-1}(f(x))| \rfloor +\log n$ 
\[ (1-o(1)) \left(1-\left(\frac{1}{n}\right)^{2^{k-j}} \right) \leq p_j \leq
1-\left(\frac{1}{n}\right)^{2^{k-j}}\,.  \]
\end{lemme}

\subsubsection{The construction of the Quantum Sampler}

Here, we reproduce the Impagliazzo and Luby construction in the quantum
setting. Most of the analysis remains the same and hence we do not repeat all the details, however we highlight the places where the analysis differs.

As before, let $f:\01^*\rightarrow \01^*$ be the candidate
quantum distributionally one-way function, fix the input size to be $n$, and define $g(x,h_k) =
(f(x), h_k,h_k(x))$. Assuming that we have a quantum inverter $I$ for
$g$, our goal is to construct a quantum sampler for $f$, namely the
following unitary
\[ \mbox{QSampler: } \ket{f(x)}\ket{0} \mapsto a_{f(x)}
\ket{f(x)}\ket{H_{f(x)}} + b_{f(x)}\ket{G_{f(x)}}, \]  
where $\frac{1}{2^n}\sum_{x\in\{0,1\}^n} a^2_{f(x)}
\geq 1 - o(1)$ and
$\ket{H_{f(x)}}~=~\frac{1}{\sqrt{|f^{-1}(f(x))|}}\sum_{x\in 
f^{-1}(f(x))}\ket{x}$. 

Similar to the classical case, we restrict ourselves to `good' $f(x)$'s.
First, we assume that for a given $f(x)$ we know the
size of the preimage $|f^{-1}(f(x))|$ and $k = \lfloor \log
|f^{-1}(f(x)) |\rfloor +O(\log n)$. 
The following unitary operations are the quantum equivalents of
picking a random universal hash function $h_k$ and a random string $r_k\in\01^k$
and are efficiently constructible:   
\begin{eqnarray*}
Q &:& \ket{k}\ket{ 0} \rightarrow \ket{k} \frac{1}{\sqrt{|H|}} \sum_{h_k}
\ket{h_k} \;\;\;,\;\;\;
B \;\;:\;\; \ket{k}\ket{ 0} \rightarrow \ket{k} \frac{1}{2^{k/2}} \sum_{r_k \in \01^k}
\ket{r_k} ,\,
\end{eqnarray*}
where $H$ is the number of possible universal hash functions $h_k : \01^n \rightarrow \01^k$\footnote{In fact, similar to the classical case one has to use a polynomial number of independent universal hash functions instead of one.}. From what follows we drop the above normalization factors. 

Let us, now, define a perfect inverter $I$ for $g$. The inverter, given an input $(f(x),h_k,h_k(x))$, such that there exists a unique $x\in  f^{-1}(f(x))$ mapped to $h_k(x)$, always returns $x$ and given an input $(f(x),h_k,s_k)$, such that there is no $x\in f^{-1}(f(x))$ mapped to $s_k$, returns an ``error'' symbol.   
\begin{eqnarray*}
I &:& \left\{ \begin{array}{lll}
\ket{f(x)}\ket{h_k}\ket{h_k(x)}\ket{0}\ket{0} &\rightarrow &
\ket{f(x)}\ket{h_k}\ket{h_k(x)}\ket{x}\ket{0}\\ 
 \ket{f(x)}\ket{h_k}\ket{s_k}\ket{0}\ket{0} & \rightarrow &
\ket{f(x)}\ket{h_k}\ket{s_k}\ket{0}\ket{1} \end{array}\right\}\,.
\end{eqnarray*}

The last register-input to $I$ acts as the ``error flag''. Note first, that by the analysis of \cite{IL89} the errors from the fact that $h_k$ may not be one-to-one are small. Also, the inverter of $g$ is not guaranteed to be perfect but only work with probability $1-O(\delta)$, but these errors are also small (i.e. inverse polynomially small). For clarity of exposition, in our description of the quantum sampler we are going to use the perfect inverter of $g$ and assume that $h_k$ is a one-to-one mapping. 

Last, recall that
$h_k$ is an efficient hash function and hence, having $\ket{h_k}$ and 
$\ket{x}$ one can efficiently compute $\ket{h_k(x)}$ and construct the
following unitary:  
\AR{
T : \ket{h_k}\ket{h_k(x)}\ket{x} & \rightarrow& \ket{h_k}\ket{0}\ket{x}\,.
}
We are now ready to define a partial quantum sampler for $f(x)$, when
we know the size of its preimage. Denote by $p_{k,f(x)}$ the
probability that the perfect inverter would return a legal output for
given values of $f(x)$ and $k$. In the following, we drop the second
subscript and have $p_k = p_{k,f(x)}$. 
\\  

\noindent
{\bf Partial Quantum Sampler PQS(f(x),k)}
\setcounter{equation}{0}
\renewcommand{\theequation}{\roman{equation}}
\begin{eqnarray}
\nonumber & &\ket{f(x)}\ket{k}\ket{0}\ket{0}\ket{0}\ket{0}\\ \nonumber \\
&\stackrel{Q_3\otimes B_4}{\rightarrow} &\ket{f(x)}\ket{k} \sum_{h_k,r_k}  \ket{h_k}\ket{r_k}\ket{0}\ket{0}\\
& \stackrel{I_{1,3,4,5,6}}{\rightarrow}  &
\sqrt{p_k}\ket{f(x)}\ket{k} \sum_{h_k, h_k(x)}
\ket{h_k}\ket{h_k(x)}\ket{x}\ket{0} + \sqrt{1-p_k}\ket{f(x)}\ket{k}\sum_{h_k, s_k}
\ket{h_k}\ket{s_k}\ket{0}\ket{1} \\
&\stackrel{T_{3,4,5}}{\rightarrow}& \sqrt{p_k}\ket{f(x)}\ket{k} \sum_{h_k}
\ket{h_k}\ket{0}\sum_{x\in f^{-1}(f(x))}\ket{x} \ket{0}+ \sqrt{1-p_k} \ket{f(x)}\ket{k}\sum_{h_k, s_k}
\ket{h_k}\ket{s_k}\ket{0}\ket{1}\\
& \stackrel{Q^{\dag}_3}{\rightarrow}& \sqrt{p_k}\ket{f(x)}\ket{k}\ket{0}\ket{0}\ket{H_{f(x)}}\ket{0} +
\sqrt{1-p_k}\ket{f(x)}\ket{k}\ket{G_{f(x),k}}\ket{1}\,,
\end{eqnarray}
where $\ket{H_{f(x)}}~=~\frac{1}{\sqrt{|f^{-1}(f(x))|}}\sum_{x\in 
f^{-1}(f(x))}\ket{x}$. In the first
step, we construct a uniform superposition of all  
possible hash functions $h_k$ and random strings $r_k\in\01^k$. In the
second step, we perform the Inverter of $g$. Assuming that the inverter is
perfect and the 
mapping $x \mapsto h_k(x)$ is truly one-to-one, then the state is exactly the one in (ii). The first term corresponds to the strings
$r_k\in\01^k$ such that $r_k = h_k(x)$ for a unique $x\in
f^{-1}(f(x))$ and this happens with probability $p_k$. The second term
corresponds to the rest of the strings. In the third step, we uncompute $h_k(x)$
and in the last step we uncompute the superposition of $h_k$. 
The final state in the perfect case consists of two terms. The first
one is $\ket{f(x)}\ket{k}\ket{H_{f(x)}}$, where the third register
contains a uniform superposition of the preimages of $f(x)$ 
and the second term denotes that the Sampler has failed (``error
flag'' register is 1). The norm of 
the first term is $p_k$, which is the probability that the inverter
outputs a legal output for the given values $k,f(x)$. 

Our partial quantum sampler imitates exactly the Impagliazzo and Luby one and
hence their analysis implies exactly that conditioned on
our sampler not failing, the actual state produced at the end is very
close to the state $\ket{f(x)}\ket{k}\ket{H_{f(x)}}$. 
Moreover, since we picked $k=\lfloor \log |f^{-1}(f(x))|\rfloor+
O(\log n)$ the norm ($p_k$) of the term $\ket{f(x)}\ket{k}\ket{H_{f(x)}}$ is
not negligible. 

Though the classical and quantum partial samplers seem identical,
there is, in fact, a difference. In the above procedure,
for superposition inputs, different values of $\ket{k}$ and
$\ket{f(x)}$ get entangled and so the naive way of implementing the
classical sampler $S(f(x))$ as a quantum circuit will fail. This can
be overcome by applying the classical procedure in a ``clean'' way \ie
garbage-free where the garbage in this case is the $\ket{k}$
register. However, since the classical procedure consists of a ``While
Loop'' (a loop with an exit command) the procedure of un-computing the
garbage is more demanding than the usual case where one deals with a
``For Loop''. To do so, instead of implementing the while loop of the
classical algorithm we prepare a weighted superposition of all $k$'s as
an ancilla register which then leads to our garbage-free quantum
sampler. 

First we construct a partial ancilla preparation circuit for
the case where the value of $k$ is known. Basically, we apply our
partial quantum sampler twice in order to ``clean'' the register that
contains $\ket{H_{f(x)}}$, while copying the ``error flag'' in
between.\\   

\noindent
{\bf Partial Ancilla Preparation, PAP(f(x),k)}
\begin{eqnarray*}
& &\ket{f(x)}\ket{k}\ket{0}\ket{0}\ket{0}\\
&\stackrel{PQS(f(x),k)}{\rightarrow}& \sqrt{p_k}\ket{f(x)}\ket{k}\ket{H_{f(x)}}\ket{0}\ket{0} +
\sqrt{1-p_k}\ket{f(x)}\ket{k}\ket{G_{f(x),k}}\ket{1}\ket{0} \\
& \stackrel{(\rm ctrl-NOT)_{4,5}}{\rightarrow}  &\sqrt{p_k}\ket{f(x)}\ket{k}\ket{H_{f(x)}}\ket{0}\ket{0} +
\sqrt{1-p_k}\ket{f(x)}\ket{k}\ket{G_{f(x),k}}\ket{1}\ket{1} \\
&\stackrel{PQS(f(x),k)^{\dag}}{\rightarrow}&\sqrt{p_k}\Big(\sqrt{p_k}\ket{f(x)}\ket{k}\ket{0}\ket{0}
+ \sqrt{1-p_k}\ket{f(x)}\ket{k}\ket{G'}\Big)\ket{0}+\\ 
&&
\sqrt{1-p_k}\Big(\sqrt{1-p_k}\ket{f(x)}\ket{k}\ket{0}\ket{0} + \sqrt{p_k}\ket{f(x)}\ket{k}\ket{G^{''}}\Big)\ket{1}\\
&=&\ket{f(x)}\ket{k}\ket{0}\ket{0}\Big(p_k\ket{0}+(1-p_k)\ket{1}\Big)+\\
&&\sqrt{p_k(1-p_k)} ( \ket{f(x)}\ket{k}\ket{G'}\ket{0}+ \ket{f(x)}\ket{k}\ket{G^{''}}\ket{1}) \, .
\end{eqnarray*}
We rewrite the transformation $PAP(f(x),k)$ by adding a flag register that is 1 when the third register is not $\ket{0}$ and also for clarity we do not write the third the fourth registers 
\[ PAP(f(x),k): \ket{f(x)}\ket{k}\ket{0}\ket{0} \mapsto 
\ket{f(x)}\ket{k}\Big(p_k\ket{0}+(1-p_k)\ket{1}\Big)\ket{0}+\ket{G_{f(x),k}}\ket{1}\,.
\]

We now describe a
circuit for the ancilla preparation when we start our algorithm for a
large value of $k$ and decrease it at each step by one. For clarity,
the quantum registers contain the values $n$ to $1$ instead of
$n+O(\log n)$ to $O(\log n)$ which are the real values for which the
Sampler is run. Furthermore, all the operations are controlled by
the ``error flag'' being the last register. \\

\noindent
{\bf  Ancilla Preparation AP(f(x))}
\begin{eqnarray*}
& &\ket{f(x)}\ket{n}\ket{0}\ket{n-1}\ket{0}\cdots\ket{1}\ket{0}\ket{0}\\
&\stackrel{PAP_{1,2,3}}{\rightarrow}&
\ket{f(x)}\ket{n}\Big(p_n\ket{0}+(1-p_n)\ket{1}\Big)\ket{n-1}\ket{0}\cdots\ket{1}\ket{0}\ket{0}+\ket{G}\ket{1}\\ 
&\stackrel{{\rm ctr}_3-{PAP}_{1,4,5}}{\rightarrow}& 
\ket{f(x)}\ket{n}p_n\ket{0}\ket{n-1}\ket{0}\cdots\ket{1}\ket{0}\ket{0} +\\
& &
\ket{f(x)}\ket{n}(1-p_n)\ket{1}\ket{n-1}\Big(p_{n-1}\ket{0}+(1-p_{n-1})\ket{1}\Big)\cdots\ket{1}\ket{0}\ket{0} +\\
& &\ket{G'}\ket{1}\\
&\stackrel{{\rm ctr}_5-{PAP}_{1,6,7}}{\rightarrow}& 
\ket{f(x)}\ket{n}p_n\ket{0}\ket{n-1}\ket{0}\cdots\ket{1}\ket{0}\ket{0}\\
&+&
\ket{f(x)}\ket{n}(1-p_n)\ket{1}\ket{n-1}p_{n-1}\ket{0}\cdots\ket{1}\ket{0}\ket{0}\\
&+& 
\ket{f(x)}\ket{n}(1-p_n)\ket{1}\ket{n-1}(1-p_{n-1})\ket{1}\ket{n-2}\Big(p_{n-2}\ket{0}+(1-p_{n-2})\ket{1}\Big)\cdots\ket{1}\ket{0}\ket{0}\\
&+&\ket{G''}\ket{1}\\
& \rightarrow&\\
&\vdots&\\
&\rightarrow&
\ket{f(x)}\ket{n}\cdots\ket{1}\sum_{j} q_j\ket{j}\ket{0} + \ket{G_f}\ket{1}\,,
\end{eqnarray*}
where $q_j=\prod_{i=1}^{j-1}(1-p_i)p_j$ is the probability that the
sampler $PQS$ succeeds at the $j$-th round and has failed on all
previous rounds. Since the registers that contain the values $n$ to $1$
are not entangled with $f(x)$ we can ignore them and have
\[ AP: \ket{f(x)}\ket{0}\ket{0} \mapsto
\ket{f(x)}\sum_{j} q_j\ket{j}\ket{0}+\ket{G_f}\ket{1}\,.  \]

Now we present the garbage-free quantum sampler
for the general case where we don't know the size of the pre-image for
a given $f(x)$. For clarity, we don't explicitly write down all the
necessary $\ket{0}$ registers in every step and also all the unitaries
are performed when the ``error flag'' is 0. \\  

\noindent
{\bf Quantum Sampler, QS(f(x))}
\begin{eqnarray*}
& &\ket{f(x)}\ket{0}\ket{0}\\
&\stackrel{AP}{\rightarrow}& \ket{f(x)}\sum_{j} q_j\ket{j}\ket{0} +
\ket{G^1_{f(x)}}\ket{1}\\ 
&\stackrel{PQS}{\rightarrow}& 
\ket{f(x)}\sum_{j}
q_j\ket{j}\Big(\sqrt{p_j}\ket{H_{f(x)}}\ket{0}+\sqrt{1-p_j}\ket{G^{2}_{f(x),j}}\ket{1}\Big)\ket{0}+\ket{G^{1}_{f(x)}}\ket{1}\\
&=& \ket{f(x)}\sum_{j} q_j\sqrt{p_j}\ket{j}\ket{H_{f(x)}}\ket{0} + \ket{G^{3}_{f(x)}}\ket{1}\\
&\stackrel{AP^{\dag}}{\rightarrow}&
\sum_{j}q_j^2\sqrt{p_j}\ket{f(x)}\ket{H_{f(x)}}\ket{0}+\ket{G^4_{f(x)}}\ket{1}\,,
\end{eqnarray*}
where the last step follows from the unitarity of $AP^\dagger$, \ie from 
\begin{eqnarray*}
\ket{f(x)}\sum_{j} q_j\ket{j}\ket{0} + \ket{G^1_{f(x)}}\ket{1}
&\stackrel{AP^\dagger}{\rightarrow}& \ket{f(x)}\ket{0}\ket{0}  \\
\ket{f(x)}\sum_{j} q_j\sqrt{p_j}\ket{j}\ket{0}
&\stackrel{AP^\dagger}{\rightarrow}& \alpha \ket{f(x)}\ket{0}\ket{0} +\beta
\ket{G}\ket{1}\,.
\end{eqnarray*} 
We conclude that $\alpha = \Big( \bra{f(x)}\sum_{j} q_j\bra{j}\bra{0} +
\bra{G^1_{f(x)}}\bra{1} \Big)\Big( \ket{f(x)}\sum_{j}
q_j\sqrt{p_j}\ket{j}\ket{0}\Big) = \sum_{j}q_j^2\sqrt{p_j}$.  

It remains to compute the success probability of the
Garbage-free Quantum Sampler, i.e to calculate the square of the sum 
$\sum_{j}q_j^2\sqrt{p_j}$. Proving that it is $1-o(1)$, then we
obtain a contradiction to $f$ being a quantum distributionally one-way
function and hence we conclude that $g$ is a quantum one-way function. 
Note that the success probability of the Impagliazzo and Luby sampler is
 $\sum_j q_j$ and Lemma \ref{pk} proves that for $j \geq k
= \lfloor \log |f^{-1}(f(x))|\rfloor + O(\log n)$ one obtains
$\sum_{j\geq k}q_j = 1-o(1)$. Here, we have a slightly more
complicated expression that can still be shown to be large. 
\begin{lemme} \label{sum}
The procedure $QS$ is a quantum sampler for $f$
with probability $1-o(1)$, i.e. $\sum_{j}q_j^2\sqrt{p_j} \geq
1-o(1)$. 
\end{lemme}

\begin{proof} 
We are going to bound this sum by showing that there exists a
particular $m$ for which the term $q_m^2\sqrt{p_m}$ is $1-o(1)$.
In order to do so, we slightly change the procedure we described above
and instead of starting from $j=n+\log n$ and decreasing $j$ at each
step by 1, we 
pick a random offset $r\in [\log \log n]$, start with $j=n+\log n +r$
and decrease $j$ at each step by $\log \log n$. Also, let $k =
\lfloor \log |f^{-1}(f(x))|\rfloor +\log n$. The values of $p_j$ for
different $j$'s can be estimated using Lemma \ref{pk} 
\[ (1-o(1)) \left(1-\left(\frac{1}{n}\right)^{2^{k-j}} \right) \leq p_j \leq
1-\left(\frac{1}{n}\right)^{2^{k-j}}\,.  \]
First, we bound the probability that the algorithm fails in all the
rounds for $j=n+\log n+r$ to $j\geq k + (1+\epsilon)\log \log n$,
where for example $\epsilon=\frac{1}{\log \log \log n}$. Note 
that at each round $j$ is decreased by $\log \log n$.  
Since $p_j$ is a decreasing function of $j$ the minimum probability
of failure is obtained for $r=0$ and is
\begin{eqnarray*}
\prod_{j=k+(1+\epsilon)\log\log n}^{n+\log n} (1-p_j) &\geq&
\prod_{j=k+(1+\epsilon)\log\log 
n}^{n+\log n} \left( \frac{1}{n} \right)^{2^{k-j}} = \prod_{\ell\geq
1}    
\left( \frac{1}{n} \right)^{2^{-(\ell+\epsilon)\log\log n}} \\
&=& 
\prod_{\ell\geq 1} \left( \frac{1}{n} \right)^{(\log
n)^{-(\ell+\epsilon)}} =  
\left( \frac{1}{n} \right)^{\sum_{\ell\geq 1} (\log
n)^{-(\ell+\epsilon)}}\\ 
& \approx &  \left( \frac{1}{n} \right)^{\frac{1}{(\log
n)^{1+\epsilon}-1}} = 1-o(1) \,.
\end{eqnarray*}
Moreover, for any $j\in [k+ \epsilon \log\log n, k+(1-\epsilon)\log
\log n]$,  we have that
\[ p_j \geq  1-\left(\frac{1}{n}\right)^{2^{-(1-\epsilon)\log \log
n}} = 1 - \left(\frac{1}{n}\right)^{(\log n)^{-(1-\epsilon)}} =   
 1 - \left(\frac{1}{2}\right)^{(\log n)^{\epsilon}} = 1-o(1) \,.
\]
Since we pick a random initial offset $r\in [1,\log\log n]$, then with 
probability $(1-2\epsilon)$ over $r$ the algorithm 
is run for an $m \in [k + \epsilon \log \log n, k+(1-\epsilon)\log
\log n]$. In this case, we have already shown that $p_m =
1-o(1)$ and, moreover, for all previous rounds 
we have $j\geq k+(1+\epsilon)\log\log n$ and hence the
probability of failure is $\prod_{j > m}(1-p_j)= 1-o(1)$. To sum up,
with probability $(1-2\epsilon)= 1-o(1)$ our algorithm is run for an
$m$ such that 
\[  \sum_{j}q_j^2\sqrt{p_j} \geq q_m^2\sqrt{p_m} =
\prod_{j > m} (1-p_j)^2 p_m^{5/2} = 1-o(1)\,, \]  
and therefore the overall success probability of the algorithm is
$1-o(1)$.   
\end{proof}

\noindent
This concludes
the proof of Theorem \ref{t-dist2oneway} and together with Theorem
\ref{t-distqs} we have  

\begin{theo} 
\label{t-mqs} Assume for a classical circuit $C$, which computes a
many-to-one function,  the corresponding 
CQS problem is hard , \ie there exists no poly($|C|$) size
quantum circuit implementing $QS_C$. Then there exists a quantum
one-way function.  
\end{theo}


\section{Statistical Zero Knowledge and quantum one-way functions}

The CQS problem has an interesting connection to the classical
complexity class of Statistical Zero Knowledge (SZK) languages:   
\TH \cite{AT02}
Any language $\mathcal{L} \in \rm{SZK}$ can be reduced to a set of
instances of the CQS problem. 
\HT

The proof is based on a reduction of the following SZK-complete
problem to a quantum sampling problem.

\DE 
\label{d-szk}\cite{SV97}
Consider two
constants $0\leq \beta < \al \leq 1$ such that $\al^2 > 
\beta$. Statistical Difference  ($SD_{\al,\beta}$) is the promise
problem of deciding for any 
two given classical circuits $C_0$ and $C_1$ whether their output
distributions are close to or far from each other, i.e. whether: 
\AR{
|| D_{C_0} - D_{C_1}|| &\geq& \al \;\;\;\;\; \mbox{or}\;\;\;\;\;||
D_{C_0} - D_{C_1}|| &\leq& \beta \,. 
 } 
\ED

It is not hard to see that the above problem can be reduced to the
problem of quantum sampling the circuits $C_0$ and $C_1$. Indeed, if one
could efficiently construct the quantum samples $\ket{C_0}$ and
$\ket{C_1}$, then, by performing a SWAP-test, one could decide whether the
two circuit distributions are close to or far from each
other. Equivalently, the above problem can be reduced to the problem
of quantum sampling the circuit $C \stackrel{\mbox{\tiny$\triangle$}}{=} C_0 \otimes C_1$, since a SWAP-test
would again decide whether the two circuit distributions are close or
far. Based on this result, we obtain the quantum analog of Ostrovsky's
result \cite{Ost91}:

\begin{theo}\label{t-szk}
Assume there exists a language  $\mcl{L} \in \rm{SZK} \setminus
\rm{AvgBQP}$, then quantum one-way functions exist.
\end{theo}

\begin{proof}
Assume $\mcl{L} \in \rm{SZK} \setminus \rm{AvgBQP}$. For every input size $n$, let $\{C^x\}_{x\in\01^n}$ be the set of classical circuits which decide $L$ via reduction to the complete language in Definition \ref{d-szk}. Denote by $m=poly(n)$ the size of the input to the circuits from this set. Since the language $\mcl{L}$ is
not in $\rm{AvgBQP}$, for any sufficiently large input size $n$, there
exists a samplable distribution ${\cal D}_n$ such that for $x \sim
{\cal D}_n$, the language $\mcl L$ can not be decided with high probability with a polynomial time quantum algorithm. Equivalently there is no polynomial
quantum algorithm that produces a quantum sample of $C^x$ for an average $x \sim 
{\cal D}_n$. We can assume this distribution to be uniform \cite{IL90}
and hence we have a uniform family of sets of circuits $\{ \{ C^x \}_{x\in\01^n} \}_{n\in {\bf N}}$, such that for any  polynomial time quantum
algorithm $Q$, any constant $\epsilon\in [0,1/2)$, and all
sufficiently large $n\in{\bf N}$  
\[ 
Q: \ket{x}\ket{0}  \mapsto c_x \ket{x}\ket{C^x} + d_x \ket{G_x} \,,
\]
with 
\[
\frac{1}{2^n}\sum_{x}  |
\inp{Q(\ket{x}\ket{0})}{\ket{x}\ket{C^x}}|^2
=\frac{1}{2^n}\sum_{x} |c_x|^2 < 1 - \epsilon\,.
\] 
We define the function $f_C:\01^* \rightarrow \01^*$ such that
$f_C: (x,y)\mapsto (x,C^x(y))$ and prove that it is a quantum one-way
function. We assume that $f$ is one-to-one otherwise from Theorem \ref{t-dist2oneway}, we can obtain the same result.   
Suppose that the function $f_C$ is not one-way, then there exists an
inverter such that
\[ I: \ket{f(x,y)}\ket{0}\ket{0} \mapsto a_{f(x,y)}
\ket{f(x,y)}\ket{x}\ket{y}+ 
b_{f(x,y)}\ket{G_{f(x,y)}}\,, \] 
or equivalently
\[ I: \ket{x}\ket{C^x(y)}\ket{0} \mapsto a_{f(x,y)} \ket{x}\ket{C^x(y)}\ket{y}+
b_{f(x,y)}\ket{G_{f(x,y)}}\,, \]   
where $\frac{1}{2^{n+m}}\sum_{x,y}a^2_{f(x,y)} \geq 1-\frac{1}{p(n)}$
(the average is taken over $x$ and $y$) and the $a_{f(x,y)}$'s are
positive. We start from a uniform 
superposition of all $y$ and use the inverter to create a 
circuit that is a good-on-average quantum sampler (similar to the
proof of Theorem \ref{t-qs}): 
\AR{
\ket{x}\frac{1}{2^{m/2}}\sum_{y}\ket{y}\ket{0}&\rightarrow_{U_f}&
\ket{x}\frac{1}{2^{m/2}}\sum_{y}\ket{y}\ket{C^x(y)}\\
&\rightarrow_{\rm{SWAP}}&
\ket{x}\frac{1}{2^{m/2}}\sum_{y}\ket{C^x(y)}\ket{y}\\
&\rightarrow_I& \ket{x}\frac{1}{2^{m/2}}\sum_{y} ( a_{f(x,y)}
\ket{C^x(y)}\ket{0}+
b_{f(x,y)}\ket{G'_{f(x,y)}} ) \equiv \ket{x}\ket{T_m} \,,
}
and hence for an average $x$
\begin{eqnarray*}
\frac{1}{2^n}\sum_{x} | \bra{x}\bra{T_m}\ket{x}\ket{C^x}|^2 & = &
\frac{1}{2^n}\sum_{x} | \frac{1}{2^{m}}\sum_{y}a_{f(x,y)}|^2 
\;\; \geq \;\; | \frac{1}{2^{n+m}}\sum_{x,y}a_{f(x,y)}|^2 \\
& \geq & | \frac{1}{2^{n+m}}\sum_{x,y}a^2_{f(x,y)}|^2 
\;\; \geq \;\;  ( 1 - \frac{1}{p(n)})^2 \;\;  \geq \;\; 1-\epsilon \,.
\end{eqnarray*}
This is a contradiction and hence the function $f_C$ is a quantum one-way. 
\end{proof}


\section{Conclusions}

In this paper we prove that the existence of any problem in SZK which
is hard-on-average for a quantum computer, implies the existence of
quantum one-way functions. Our proofs go through the problem of
quantum sampling.    
Aharonov and Ta-Shma cast many important problems as quantum sampling
problems and described a possible way for attacking them. It is,
hence, very interesting to investigate the 
real hardness of quantum sampling. We already know that if
$\rm{SZK}\not\subseteq \rm{AvgBQP}$ then there exist hard instances of quantum
sampling. Under what other assumptions can one prove the existence of
hard instances of the CQS problem and consequently quantum one-way functions? 

Furthermore, we saw that our candidate one-way problems include some
of the most 
notorious problems in quantum computing, like Graph Non-Isomorphism and
approximate Closest Lattice Vector problem. Could we construct one-way
functions from other problems, such as the hidden subgroup problem in
the dihedral or other non-abelian groups? 

Last, Watrous \cite{W05} proved that computational zero
knowledge for NP is implied by the existence of quantum one-way
permutations. What other implications does the existence of
quantum one-way functions have? 

\section*{Acknowledgements}
We wish to thank Oded Regev, Ben Reichardt and Alain Tapp for useful discussions and
Andrej Bogdanov for sharing with us his notes on \cite{IL89}. 
EK was partially supported by the ARDA, MITACS, ORDCF, and CFI projects during her stay at University of Waterloo where this work was begun. IK gratefully acknowledges professor Peter w. Shor who has
supported his research from a scholarly allowance provided from his
appointment as the holder of the Henry Adams Morss and Henry Adams
Morss, Jr. Professorship and from NSF CCF-0431787.


\end{document}